\documentclass[aip,rsi,reprint,graphicx]{revtex4-1} 
\usepackage{graphicx}
\usepackage{dcolumn}
\usepackage{booktabs}
\usepackage{bm}
\usepackage{amsmath}
\usepackage{amssymb}
\usepackage{latexsym}
\usepackage{epsfig}
\usepackage{amsbsy}
\usepackage{array}
\usepackage{amssymb}
\usepackage{setspace}
\usepackage{ulem}
\usepackage[colorlinks, linkcolor=blue, anchorcolor=blue, citecolor=blue, urlcolor=blue]{hyperref}
\usepackage{underscore} 
\draft 

\begin{document}

\title{Portable Microwave Frequency Dissemination in Free Space and Implications on Ground-to-Satellite Synchronization}

\author{J. Miao}
\affiliation{Joint Institute for Measurement Science, Department of Precision Instrument, Tsinghua University, Beijing 100084, China}
\affiliation{Department of Physics, Tsinghua University, Beijing 100084, China}

\author{B.~Wang}
\email[Electronic mail: ]{bo.wang@tsinghua.edu.cn}
\affiliation{Joint Institute for Measurement Science, Department of Precision Instrument, Tsinghua University, Beijing 100084, China}
\affiliation{State Key Lab of Precision Measurement Technology and Instrument, Tsinghua University, Beijing 100084, China}

\author{Y.~Bai}
\affiliation{Joint Institute for Measurement Science, Department of Precision Instrument, Tsinghua University, Beijing 100084, China}
\affiliation{Department of Physics, Tsinghua University, Beijing 100084, China}

\author{Y.~B.~Yuan}
\affiliation{Joint Institute for Measurement Science, Department of Precision Instrument, Tsinghua University, Beijing 100084, China}
\affiliation{Department of Physics, Tsinghua University, Beijing 100084, China}

\author{C.~Gao}
\affiliation{Joint Institute for Measurement Science, Department of Precision Instrument, Tsinghua University, Beijing 100084, China}
\affiliation{State Key Lab of Precision Measurement Technology and Instrument, Tsinghua University, Beijing 100084, China}

\author{L.~J.~Wang}
\email[Electronic mail: ]{lwan@tsinghua.edu.cn}
\affiliation{Joint Institute for Measurement Science, Department of Precision Instrument, Tsinghua University, Beijing 100084, China}
\affiliation{Department of Physics, Tsinghua University, Beijing 100084, China}
\affiliation{State Key Lab of Precision Measurement Technology and Instrument, Tsinghua University, Beijing 100084, China}
\affiliation{National Institute of Metrology, Beijing 100013, China}

\date{Received 15 March 2015; accepted 30 April 2015; published online 18 May 2015}

\begin{abstract}
Frequency dissemination and synchronization in free space plays an important role in global navigation satellite system, radio astronomy and synthetic aperture radar. In this paper, we demonstrated a portable radio frequency dissemination scheme via free space using microwave antennas. The setup has a good environment adaptability and high dissemination stability. The frequency signal was disseminated at different distances ranging from 10 to 640\,m with a fixed 10\,Hz locking bandwidth, and the scaling law of dissemination stability on distance and averaging time was discussed. The preliminary extrapolation shows that the dissemination stability may reach $1\times10^{-12}$/s in ground-to-satellite synchronization, which far exceeds all present methods, and is worthy for further study.[\href{http://dx.doi.org/10.1063/1.4921001}{http://dx.doi.org/10.1063/1.4921001}]
\end{abstract}


\maketitle

\section{Introduction}

In recent years, high-precision time and frequency signals dissemination via optical fiber has shown rapid progress.\cite{Ma1994,Ye2003,Levine2008,Warrington2012} Dissemination schemes for different topological structures, from point-to-point, multi-access, cascade to branching structure have been experimentally demonstrated.\cite{Kumagai2009,Lopez2010,Predehl2012,Wang2012,Gao2012,Bai2013,Grosche2014} It is useful for positioning, navigation, timing, radio astronomy (such as Very Long Baseline Interferometry and Square Kilometer Array) and measurement of fundamental constants. However, the fiber-based dissemination schemes can only realize two-dimensional frequency synchronization, and the disseminated signals are confined in the fiber network. Consequently, for portable and three-dimensional applications or any cases with no fiber sources, free space time and frequency transfer is essential and is in great need for improvement. Most of the related experiments made use of free space laser.\cite{Sprenger2009,Djerroud2010,Nie2012,Giorgetta2013} For example, F.~R.~Giorgetta {\it et al}. developed an optical two-way time and frequency transfer (TWTFT) method which can exchange optical pulse between coherent frequency combs, and they achieved a residual instability below $1\times10^{-18}$ at 1000\,s.\cite{Giorgetta2013} However, there are some limitations in optical time and frequency dissemination via free space. The transmission quality is highly affected by weather conditions, and it is hard to maintain an effective transfer link on a large scale because of the small beam divergence, complex operations and environmentally sensitive equipment. As a result, optical transfer is not capable of all-weather dissemination.

By compensating phase noise actively,
 we have successfully disseminated the L-band frequency signal between RF antennas in free space with a stability of $3\times10^{-13}$/s and $4\times10^{-17}$/day.\cite{Miao2013} To study potential applications of phase compensated radio frequency dissemination scheme, especially the application in the case of ground-to-satellite synchronization, we improved the experimental approach and selected an appropriate field test site. The links are set with a fixed 10\,Hz locking bandwidth to simulate actual ground-to-satellite communication. The dissemination stability (denoted as $\sigma$) of the links from 0.5\,s to 1000\,s averaging time (denoted as $\tau$) at different distances (denoted as $d$) ranging from 10\,m to 640\,m was measured in steady nocturnal environment. Using a binary regression model, we explored the relationship between $\tau$, $d$ and $\sigma$. The regression result shows that at a certain averaging time,  $\lg(\sigma)$ degrades linearly as $\lg(d)$ increases. If this scaling law still applies in longer distance, the dissemination stability is expected to be better than 1$\times$10$^{-12}$/s in a ground-to-satellite synchronization link, which is superior to traditional two-way satellite time and frequency transfer (TWSTFT) and synchronization methods based on Global Positioning System (GPS).\cite{Bauch2006}

The details of experimental setup are presented in Sec.~\ref{ii}, and the experiment results and data analysis are presented in Sec.~\ref{iii}.

\section{System Description}
\label{ii}
\begin{figure}
	\centering
	\includegraphics[width=85mm]{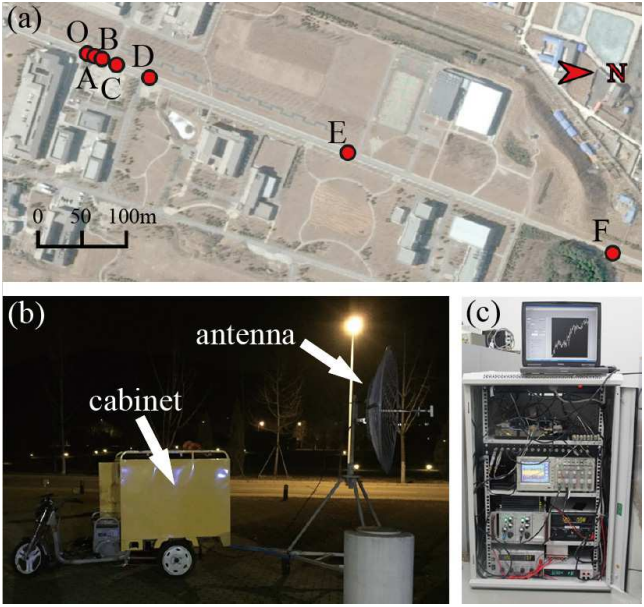}
	\caption{(a) Satellite map of the experimental site. The local terminal was placed at point $O$, and the remote terminal moved along the testing baseline from point $A$ to point $F$; $A$, $B$, $C$, $D$, $E$, $F$ are 10\,m, 20\,m, 40\,m, 80\,m, 320\,m and 640\,m away from point $O$, respectively. (b) The remote Terminal and its antenna under test. (c) The rack mount chassis inside the cabinet. \label{device}}
\end{figure}

In our previous experiment,\cite{Miao2013} due to the restriction of the experimental site, the local terminal and the remote terminal were located in the same room, and the antennas were placed on the roof top of Department of Precision Instrument Building at Tsinghua University. Consequently, we had to connect each antenna to local terminal or remote terminal with RF coaxial cable of tens-of-meters length, which introduced undesired phase noise, and restricted longer transfer distance. Recently, we built a vehicle-mounted portable experimental system to replace fixed equipment. All units of local or remote terminal are integrated in a 19-in. rack mount chassis with a height of 900\,mm, which can be placed in cabinet and work with ordinary AC power supply. A Parabolic dish antenna with a diameter of 1.5\,m, mounted on a tribrach with wheels, is connected to local or remote terminal via a 5-m-long coaxial cable. Then the tribrach is hitched to a vehicle for mobility. A series of experiments were conducted at the Changping site of the National Institute of Metrology (NIM-Changping) on the outskirts of Beijing between November 2014 to January 2015. Figure~\ref{device} (a) is the map of the experimental site, Fig.~\ref{device} (b) shows the cabinet of the remote terminal and the antenna under test, and Fig.~\ref{device} (c) shows the rack mount chassis of the remote terminal, which was placed in the cabinet during the experiments.

\begin{figure*}
	\centering
	\includegraphics[width=110mm]{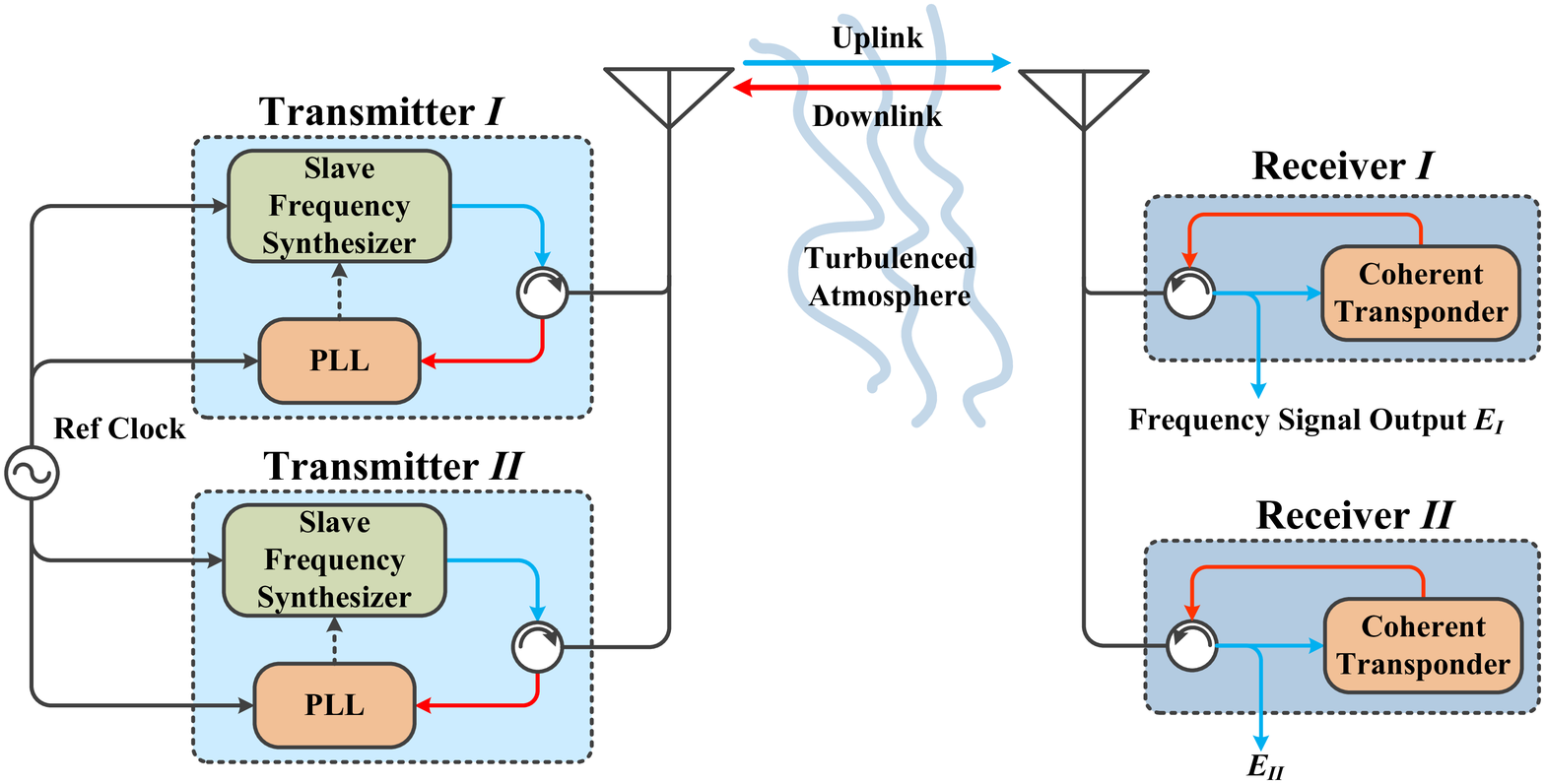}
	\caption{Schematic of the portable dissemination system.\label{principle}}
\end{figure*}

The experimental scheme diagram is shown in Fig.~\ref{principle}. There are two independent links, which can disseminate different frequency signals phase locked to a same reference clock and recover the reference frequency at the remote terminal, respectively. Link $I$ contains Transmitter $I$ and Receiver $I$ while Link $II$ contains Transmitter $II$ and Receiver $II$. Take Link $I$ for example, at local terminal the 100\,MHz reference clock generates a signal $E_r=V_r \cos(\omega_r t+\phi_r)$. In the slave frequency synthesizer there is an oscillator generating $E_0=V_0 \cos(\omega t+\phi_0)$, and there are several multipliers to generate frequency signals coherent to $E_0$. Here, $\omega$ is nominally 100\,MHz and $\phi_0$ is the initial phase of the oscillator. When the uplink signal reaches Receiver $I$, it is coherent to the output frequency signal $E_I=V_I \cos(\omega t+\phi_0+\phi_p)$, where $\phi_p$ is a variable standing for the phase (with noise) accumulated during transmission. Through a coherent transponder in Receiver $I$, the signal is returned to local terminal and is coherent to $E_R=V_R \cos(\omega t+\phi_0+2\phi_p)$, after a round-trip. By a two-stage mixing at the transmitter, an error signal $E_e=V_e\cos[(\omega_r-\omega)t+(\phi_0+\phi_p-\phi_r)]$ is generated, which is fed to a phase-locked loop (PLL), making $\omega=\omega_r$ and $\phi_0+\phi_p=\phi_r+constant$ when the PLL is locked. Hence, $E_I$ is phase locked to the reference signal, The same applies to Link $II$ ($E_{II}$). The phase difference between $E_I$ and $E_{II}$ stands for $\sqrt 2$ times the dissemination stability of each link. For convenience, we ignore the factor $\sqrt 2$ in the following discussion. Detailed information about the structure of transmitters and receivers is described in our previous paper.\cite{Miao2013}

Frequency signals suffer turbulence and temperature fluctuation during free space dissemination, which mainly result in changes of refractive index. A.~N.~Kolmogorov deduced a structure function of the velocity field in turbulent flows at very large Reynolds numbers:\cite{Kolmogorov1941,Kolmogorov1962,Sinclair2014}

\begin{equation}
D_{v}^{2}(r)=C_{v}^{2}r^{2/3},\;l_{0} \leqslant r \leqslant L_{0},
\end{equation}
where $C_{v}^{2}$ is the velocity structure constant; $l_{0}$ is inner scale and below which viscous dissipation is negligible; $L_{0}$ is the outer scale over which no kinetic energy is injected. This function is often called ``2/3 power law."

Accordingly, structure function of refractive index is
\begin{equation}
D_{n}^{2}(r)=C_{n}^{2}r^{2/3},\;l_{0} \leqslant r \leqslant L_{0},
\end{equation}
where $C_{n}^{2}$ is the structure constant of refractive index. Then after a Fourier transform, we get
\begin{equation}
\Phi_{n}(\kappa)=0.033C_{n}^{2}\kappa^{-11/3},\;2\pi/L_{0} \leqslant \kappa \leqslant 2\pi/l_{0},
\end{equation}
where $\kappa$ is the spatial frequency, $\Phi_{n}$ is the power spectral density of refractive index.\cite{Tatarski1961}

The frequency stability of phase compensated round-trip dissemination scheme may obey the power law similar to $D_{n}^{2}(r)$. To test this hypothesis and preliminarily study the feasibility for ground-to-satellite applications, we performed a field test using the 700-m-long baseline in NIM-Changping. The local terminal was fixed at point $O$, and we moved the remote terminal to measure the dissemination stability (Allan deviation, ADEV) at different distances respectively. Note that even perfect phase noise cancellation with time delay in round-trip brings imperfect phase compensation in one-way transfer.\cite{Newbury2007,Williams2008} A round-trip communication takes 0.15\,s between ground and medium-earth orbit satellite (MEO) and takes 0.25\,s between ground and geostationary satellite (GEO). For these occasions, locking bandwidth of the PLL is as low as several hertz.\cite{Gardner2005} To simulate the ground-to-satellite frequency dissemination with large time delay and fixed low locking bandwidth, we set the PLL locking bandwidth to 10\,Hz.

The outdoor temperature varied from $-$12\,$^{\circ}$C to 20\,$^{\circ}$C during the experiments. Low temperature indeed brought trouble to the equipment at the beginning. It caused some devices working improperly, for example, the frequent frequency jumps of frequency synthesizers and oscillators. Then we pasted foam heat-insulation layer on the inner wall of each cabin in 1.4\,m$\times$1\,m$\times$1\,m size, making sure that all electronic components work properly. However, we can not control the temperature in the cabinet, so long-term (more than 500\,s) frequency stability does not decline as averaging time increases due to long-term relatively large temperature fluctuations in the cabinet.

\section{Modeling of Scaling Law and Data analysis}
\label{iii}
\begin{figure}
	\centering
	\includegraphics[width=85mm]{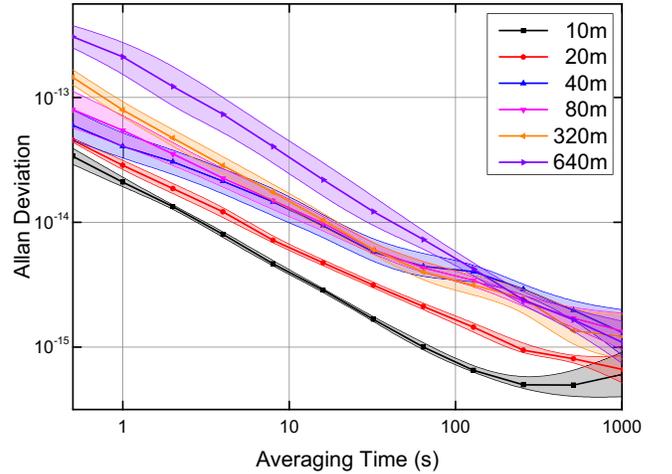}
	\caption{Measured frequency stability of the dissemination system with PLLs closed in steady nocturnal environment. Repeated measurement results at each distance are represented as the shaded area and the mean values are represented as solid lines within shaded areas.\label{ADEV}}
\end{figure}

Figure~\ref{ADEV} shows the measured frequency dissemination stabilities of the experiment system at different distances in steady nocturnal environment. At each distance, we measure repeatedly to reduce stochastic error. In the shortest  distance (10\,m) the stability is $2.1\times 10^{-14}$ at 1\,s and declines to $5.0\times 10^{-16}$ at 256\,s; in the longest distance (640\,m) the stability is $2.1\times 10^{-13}$ at 1\,s and declines to $2.4\times 10^{-15}$ at 256\,s. The frequency stability of all the distances is limited to around $1\times 10^{-15}$ after 1000\,s owing to temperature fluctuation in the cabinet.

\begin{figure}
	\centering
	\includegraphics[width=85mm]{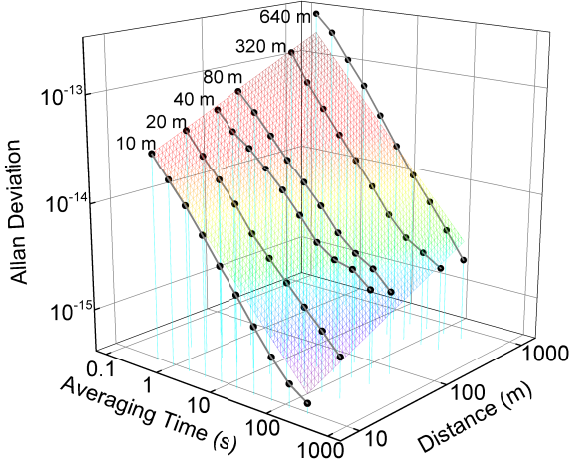}
	\caption{Fitted surface plot of distance (d), averaging time ($\tau$) and dissemination stability ($\sigma$) in logarithmic coordinate.The black dots stands for the measured data and the plain mesh in false-color grids stands for the regression equation.\label{3D}}
\end{figure}

The dissemination stability may degrade with distance at $d^{b}$ just like the Kolmogorov scaling law, here $b$ is a constant. To verify this hypothesis, we establish a binary linear regression model, supposing that
\begin{equation}\label{binary}
\sigma(\tau, d)=10^{z}\tau^{a}d^{b},
\end{equation}
i.e.
\begin{equation}
\lg(\sigma)=a\cdot \lg(\tau)+b\cdot \lg(d)+z,
\end{equation}
where $z$ and $a$ are constants. The regression result is based on 60 data points, ranging from 0.5\,s to 256\,s on the time scale and from 10\,m to 640\,m on the distance scale, shown in Table~\ref{result}. Here, SE stands for standard error. We obtain $b=0.422$, $a=-0.646$, with the 95\% confidence intervals of $b$ and $a$ being 0.369 $\sim$ 0.475 and $-0.685 \sim -0.607$, respectively. This indicates that the dissemination stability declines over averaging time at approximate $\tau^{-0.65}$ and degrades with distance at approximate $d^{0.42}$. Fig.~\ref{3D} intuitively illustrates the fitting result. The P-value is 0.000, and the adjusted R-Square for the whole equation is 0.958, showing great statistical significance. Table~\ref{PC} displays a strong positive correlation between $\lg(\sigma)$ and $\lg(d)$, and a strong negative correlation between $\lg(\sigma)$ and $\lg(\tau)$. 

\tabcolsep 0.2pt
\begin{table}[h]
	\caption{Coefficients of parameters in binary linear regression}
	\begin{center}
		\begin{tabular*}{\linewidth}{@{\extracolsep{\fill}} c   r   c   r  r }\toprule[1.5pt]
			&   &   & \multicolumn{2}{c}{95.0\% Confidence Interval}\\ \cmidrule[0.4pt]{4-5}
			& Value & SE & Lower bound & Upper bound \\\hline
			\quad $z$ \quad\qquad & $-$14.072 	 & 0.056 & $-$14.184 & $-$13.960 \\
			\quad 	$b$ \quad\qquad  & 0.422 & 0.026 & 0.369 & 0.475 \\
			
			\quad 	$a$ \quad\qquad  & $-$0.646 & 0.019 & $-$0.685 & $-$0.607 \\\bottomrule[1.5pt]
		\end{tabular*}
	\end{center} \label{result}
\end{table}

 \tabcolsep 0.2pt
 \begin{table}[h]
 	\caption{Correlations of regressors in binary linear regression}
 	\begin{center}
 		\begin{tabular*}{\linewidth}{@{\extracolsep{\fill}} l  r   r    r }\toprule[1.5pt]
 			Pearson Correlation   & $\lg(\sigma)$ & $\lg(d)$ & $\lg(\tau)$ \\\midrule[1pt]
 			$\lg(\sigma)$ & 1.000 	 & 0.425 & $-$0.883 \\
 			$\lg(d)$ & 0.425 & 1.000 & 0.000 \\
 			
 			$\lg(\tau)$ & $-$0.883 & 0.000 & 1.000  \\\bottomrule[1.5pt]
 		\end{tabular*}
 	\end{center} \label{PC}
 \end{table}
 
In order to explore the relationship between $\lg(\sigma)$ and $\lg(\tau)$ more clearly, we study the scaling law on certain averaging times. In this case the regression equation is simplified into a unary linear equation 
\begin{equation}
\sigma(d)=10^{Z}d^{b},
\end{equation}
where $10^{Z}=10^{z}\tau^{a}$ with a given $\tau$. Table~\ref{seprate} shows the regression results at different averaging times. CI stands for confidence interval and R stands for correlation coefficient. As $\tau$ increases from 0.5\,s to 256\,s, parameter $b$ decreases from approximate 1/2 to 1/3, the width of $b$'s 95\% confidence interval widens from 0.25 to 0.8, and the correlation coefficient also decreases. If temperature of the out-of-loop devices and the measurement system is well controlled (experimental records indicate significant correlation between phase fluctuation and temperature fluctuation with 4\,$\sim$\,6\,ps/$^\circ$C), the difference of $b$'s regression value between short-term stability and long-term stability should be significantly narrowed and with less apparent drift.

\tabcolsep 0.2pt
\begin{table}[h]
	\caption{Parameters in unary linear regression}
	\begin{center}
		\begin{tabular*}{\linewidth}{@{\extracolsep{\fill}}c c c  r  c  }\toprule[1.5pt]
			$\tau$ / s & $Z$ & $b$ & 95.0\% CI of $b$  & R \\\hline
			
			0.5 & $-$14.004 & 0.497 & 0.370 $\sim$ 0.625 & 0.984 \\
			1 & $-$14.198 & 0.497 & 0.319 $\sim$ 0.674 & 0.968 \\
			
			2 & $-$14.332 & 0.465 & 0.285 $\sim$ 0.645 & 0.963 \\
			
			4 & $-$14.508 & 0.452 & 0.255 $\sim$ 0.649 & 0.954 \\
			
			8 & $-$14.702 & 0.441 & 0.230 $\sim$ 0.653 & 0.945  \\
			
			16 & $-$14.852 & 0.410 & 0.189 $\sim$ 0.630 & 0.932 \\
			
			32 & $-$15.030 & 0.387 & 0.148 $\sim$ 0.625 & 0.914 \\
			
			64 & $-$15.189 & 0.377 & 0.078 $\sim$ 0.676 & 0.868 \\
			
			128 & $-$15.306 & 0.365 & $-$0.027 $\sim$ 0.756 & 0.791 \\
			
			256 & $-$15.406 & 0.333 & $-$0.065 $\sim$ 0.730 & 0.758 \\
			\bottomrule[1.5pt]
		\end{tabular*}
	\end{center} \label{seprate}
\end{table}

Suppose the atmosphere is homogeneous with same density at sea level,  the height is called scale height for the atmosphere,\cite{Ametsoc} which is expressed as

\begin{equation}
H=\dfrac{k_B T}{mg},
\end{equation}
here $k_B$ is the Boltzmann constant ($1.38\times10^{-23}$\,J/K), $T$ is the temperature of atmosphere in kelvins and we can take it as 300\,K, $m$ is the mean molecular mass of air ($4.8\times10^{-26}$ kg), $g$ is the acceleration due to gravity (9.8\,m/s$^{2}$), thus $H$ = 8800\,m. The equivalent thickness of the atmosphere in the path link between ground and satellite varies with the zenith angle, and $H$ is the Minimum. For example, the equivalent thickness of the atmosphere from a ground station in Beijing (approximately located at $40^{\circ}$N, 116$^{\circ}$E) to a GEO satellite at the same longitude is about 13000\,m. We just take a preliminary extrapolation, substitute parameters $d$=13000\,m, $a$=$-$2/3, $b$=1/2, $z$=$-$14 into Eq (\ref{binary}), a stability of $1.1\times10^{-12}$/s, $5.8\times10^{-16}$/d is estimated. The stability of a typical Cesium atomic clock (Microsemi 5071A) is $5\times10^{-12}$/s, $3\times10^{-14}$/d). Consequently, the dissemination scheme is capable of transferring frequency of a commercial atomic clock without downgrade its stability. However, as a preliminary study, we have not taken into account the path loss of practical ground-to-satellite link. The free space path loss from ground to a GEO satellite (such as Inmarsat) in L band is about 190\,dB according to Friis transmission equation, and the received downlink signal power is about $-$80\,dBm if the antenna gain is 30\,dB.\cite{Inmarsat} The signal dissemination and processing issues in this case needs further study in the future.

\section{Conclusion}

In summary, we demonstrated a continuous, portable radio frequency dissemination scheme via free space. It is easy to set up with high mobility, and it can be unattended if automatic control unit is added to the system. This flexibility gives it a great potential in three-dimensional regional or even global time and frequency network and in many scientific and engineering applications, such as clock-based geodesy and synthetic aperture radar (SAR). Besides, a preliminary study indicates that the relationship between its dissemination stability and transfer distance follows a scaling law, and a stability of $1\times10^{-12}$/s, $6\times10^{-16}$/d is promising for ground-to-satellite synchronization. However, long-term frequency stability is limited to around $1\times10^{-15}$ within km-distance mainly due to temperature drift in the experimental devices. Consequently, more rigorous experiments focusing on longer distance, longer time with fine thermal control are worthy and meaningful to verify this possibility.

\begin{acknowledgments}
This work is supported by the National Key Scientific Instrument and Equipment Development Projects (No.~2013YQ09094303) and the Beijing Higher Education Young Elite Teacher Project (No.~YETP0088).

\end{acknowledgments}


\end{document}